\documentclass[a4paper]{article}
\usepackage[utf8]{inputenc}
\usepackage{csquotes} 
\usepackage{graphicx}
\usepackage[margin=1.28in]{geometry}
\usepackage{hyperref}
\usepackage[numbers]{natbib}

\newcommand{\autocite}[1]{\cite{#1}}
\newcommand{\textcite}[1]{\cite{#1}}

\title{Inscriptis - A Python-based HTML to text conversion library optimized for knowledge extraction from the Web}
\author{Albert Weichselbraun}
\date{Swiss Institute for Information Science\\
University of Applied Sciences of the Grisons\\
Pulvermühlestrasse 57, Chur, Switzerland}

\begin{document}

\maketitle

\section{Summary}\label{summary}

\texttt{Inscriptis} provides a library, command line client and Web
service for converting HTML to plain text.

Its development has been triggered by the need to obtain accurate text
representations for knowledge extraction tasks that preserve the spatial
alignment of text without drawing upon heavyweight, browser-based
solutions such as Selenium \cite{selenium}. In contrast to
existing software packages such as HTML2text \cite{html2text}, jusText
\cite{justext} and Lynx \cite{lynx}, \texttt{Inscriptis}

\begin{enumerate}
\def\labelenumi{\arabic{enumi}.}
\item
  provides a layout-aware conversion of HTML that more closely resembles
  the rendering obtained from standard Web browsers and, therefore,
  better preserves the spatial arrangement of text elements.
  \texttt{Inscriptis} excels in terms of conversion quality, since it
  correctly converts complex HTML constructs such as nested tables and
  also interprets a subset of HTML (e.g., \texttt{align},
  \texttt{valign}) and CSS (e.g., \texttt{display},
  \texttt{white-space}, \texttt{margin-top}, \texttt{vertical-align},
  etc.) attributes that determine the text alignment.
\item
  supports annotation rules, i.e., user-provided mappings that allow for
  annotating the extracted text based on structural and semantic
  information encoded in HTML tags and attributes used for controlling
  structure and layout in the original HTML document.
\end{enumerate}

These unique features ensure that downstream knowledge extraction
components can operate on accurate text representations, and may even
use information on the semantics and structure of the original HTML
document, if annotation support has been enabled.

\section{Statement of need}\label{statement-of-need}

Research in a growing number of scientific disciplines relies upon Web
content. Li et al. \cite{li_effect_2014}, for instance, studied the impact of
company-specific News coverage on stock prices, in medicine and
pharmacovigilance social media listening plays an important role in
gathering insights into patient needs and the monitoring of adverse drug
effects \cite{convertino_usefulness_2018}, and communication sciences analyze
media coverage to obtain information on the perception and framing of
issues as well as on the rise and fall of topics within News and social
media \cite{scharl_semantic_2017,weichselbraun_adapting_2021}.

Computer science focuses on analyzing content by applying knowledge
extraction techniques such as entity recognition \cite{fu_spanner_2021} to
automatically identify entities (e.g., persons, organizations,
locations, products, etc.) within text documents, entity linking \cite{ding_jel_2021} to link these entities to knowledge bases such as Wikidata
and DBPedia, and sentiment analysis to automatically assess sentiment
polarity (i.e., positive versus negative coverage) and emotions
expressed towards these entities \cite{wang_review_2020}.

Most knowledge extraction methods operate on text and, therefore,
require an accurate conversion of HTML content which also preserves the
spatial alignment between text elements. This is particularly true for
methods drawing upon algorithms which directly or indirectly leverage
information on the proximity between terms, such as word embeddings
\cite{mikolov_distributed_2013,pennington_glove:_2014}, language models \cite{reis_transformers_2021}, sentiment analysis which often also considers the
distance between target and sentiment terms, and automatic keyword and
phrase extraction techniques.

Despite this need from within the research community, many standard HTML
to text conversion techniques are not layout aware, yielding text
representations that fail to preserve the text's spatial properties, as
illustrated below:

\begin{figure}
\centering
\includegraphics[width=\linewidth]{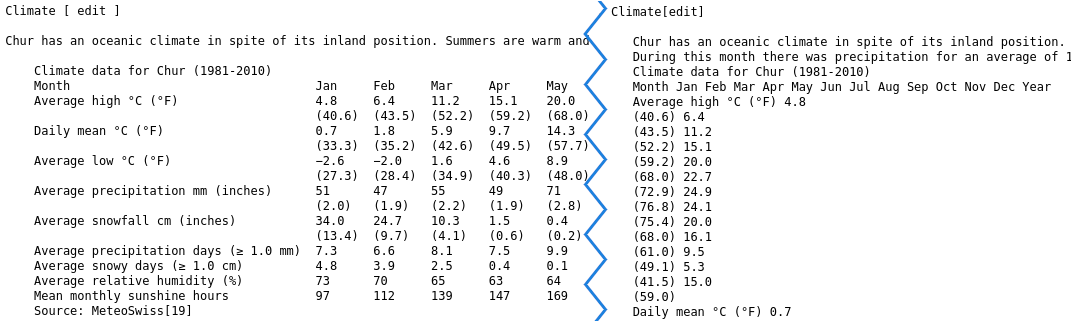}
\caption{Text representation of a table from Wikipedia computed by
\texttt{Inscriptis} (left) and Lynx (right). Lynx fails to correctly
interpret the table and, therefore, does not properly align the
temperature values.}
\end{figure}

Consequently, even popular resources extensively used in literature
suffer from such shortcomings. The text representations provided with
the Common Crawl corpus\footnote{https://commoncrawl.org/}, for
instance, have been generated with a custom utility \cite{ia-commons} which at the time of writing did not consider any layout
information. Datasets such as CCAligned \cite{el-kishky_ccaligned_2020}
multilingual C4 which has been used for training the mT5 language model
\cite{xue_mt5_2021}, and OSCAR \cite{suarez_asynchronous_2019} are based on subsets
of the Common Crawl corpus \cite{caswell_quality_2021}.

Even worse, some tutorials suggest the use of software libraries such as
Beautiful Soup \cite{beautifulsoup}, lxml \cite{lxml} and
Cheerio \cite{cheerio} for converting HTML. Since these
libraries have been designed with a different use case in mind, they are
only well-suited for scraping textual content. Once they encounter HTML
constructs such as lists and tables, these libraries are likely to
return artifacts (e.g., concatenated words), since they do not interpret
HTML semantics. The creators of the Cheerio library even warn their
users, by explicitly stating that it is not well-suited for emulating
Web browsers.

Specialized conversion tools such as HTML2Text perform considerably
better but often fail for more complex Web pages. Researchers sometimes
even draw upon text-based Web browsers such as Lynx to obtain more
accurate representations of HTML pages. These tools are complemented by
content extraction software such as jusText \cite{justext}, dragnet
\cite{peters_content_2013}, TextSweeper \cite{lang_textsweeper_2012} and boilerpy3
\cite{boilerpy3} which do not consider the page layout but rather aim at
extracting the relevant content only, and approaches that are optimized
for certain kinds of Web pages like Harvest \cite{weichselbraun_harvest_2020}
for Web forums.

\begin{figure}
\centering
\includegraphics[width=\linewidth]{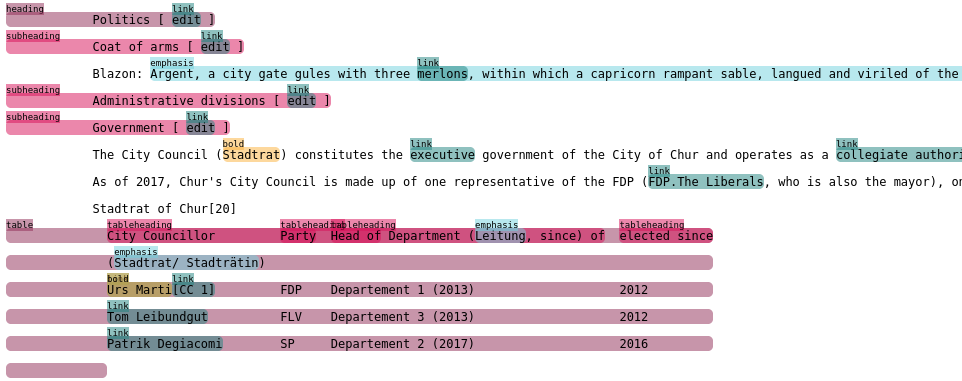}
\caption{Annotations extracted from the Wikipedia entry for Chur that
have been exported to HTML using the \texttt{-\/-postprocessor\ html}
command line option.\label{fig:annotations}}
\end{figure}

\texttt{Inscriptis}, in contrast, not only correctly renders more
complex websites but also offers the option to preserve parts of the
original HTML document's semantics (e.g., information on headings,
emphasized text, tables, etc.) by complementing the extracted text with
annotations obtained from the document. Figure~\ref{fig:annotations}
provides an example of annotations extracted from a Wikipedia page.
These annotations can be useful for

\begin{itemize}
\item
  providing downstream knowledge extraction components with additional
  information that may be leveraged to improve their respective
  performance. Text summarization techniques, for instance, can put a
  stronger emphasis on paragraphs that contain bold and italic text, and
  sentiment analysis may consider this information in addition to
  textual clues such as uppercase text.
\item
  assisting manual document annotation processes (e.g., for qualitative
  analysis or gold standard creation). \texttt{Inscriptis} supports
  multiple export formats such as XML, annotated HTML and the JSONL
  format that is used by the open source annotation tool
  doccano\footnote{Please note that doccano currently does not support
    overlapping annotations and, therefore, cannot import files
    containing overlapping annotations.} \cite{doccano}.
  Support for further annotation formats can be easily added by
  implementing custom annotation post-processors.
\item
  enabling the use of \texttt{Inscriptis} for tasks such as content
  extraction (i.e., extract task-specific relevant content from a Web
  page) which rely on information on the HTML document's structure.
\end{itemize}

In conclusion, \texttt{Inscriptis} provides knowledge extraction
components with high quality text representations of HTML documents.
Since its first public release in March 2016, \texttt{Inscriptis} has
been downloaded over 135,000 times from the Python Package Index
(PyPI)\footnote{Source: https://pepy.tech/project/inscriptis}, has
proven its capabilities in national and European research projects, and
has been integrated into commercial products such as the
\href{https://www.weblyzard.com/visual-analytics-dashboard/}{webLyzard
Web Intelligence and Visual Analytics Platform}.

\section{Mentions}\label{mentions}

The following research projects use \texttt{Inscriptis} within their
knowledge extraction pipelines:

\begin{itemize}
\item
  \href{https://www.fhgr.ch/CareerCoach}{CareerCoach}: ``Automatic
  Knowledge Extraction and Recommender Systems for Personalized Re- and
  Upskilling suggestions'' funded by Innosuisse.
\item
  \href{https://www.fhgr.ch/Job-Cockpit}{Job Cockpit}: ``Web analytics,
  data enrichment and predictive analysis for improved recruitment and
  career management processes'' funded by Innosuisse
\item
  \href{https://www.epoch-project.eu}{EPOCH project} funded by the
  Austrian Federal Ministry for Climate Action, Environment, Energy,
  Mobility and Technology (BMK) via the ICT of the Future Program.
\item
  \href{https://www.fhgr.ch/medmon}{MedMon}: ``Monitoring of Internet
  Resources for Pharmaceutical Research and Development'' funded by
  Innosuisse.
\item
  \href{https://www.retv-project.eu}{ReTV project} funded by the
  European Union's Horizon 2020 Research and Innovation Programme.
\end{itemize}

\section{Acknowledgements}\label{acknowledgements}

Work on \texttt{Inscriptis} has been conducted within the MedMon, Job
Cockpit and CareerCoach projects funded by Innosuisse.

\bibliographystyle{plain}
\bibliography{paper.bib}

\begin{thebibliography}{10}

\bibitem{lxml}
Stefan Behnel, Martijn Faassen, Ian Bicking, Holger Joukl, Simon Sapin,
  Marc-Antoine Parent, Olivier Grisel, Kasimier Buchcik, Florian Wagner, Emil
  Kroymann, Paul Everitt, Victor Ng, Robert Kern, Andreas Pakulat, David
  Sankel, Marcin Kasperski, Sidnei da~Silva, and Pascal Oberndörfer.
\newblock lxml - processing xml and html with python, 2021.
\newblock Accessed: 2021-09-02.

\bibitem{justext}
Michal Belica.
\newblock justext - heuristic based boilerplate removal tool, 2021.
\newblock Accessed: 2021-09-02.

\bibitem{caswell_quality_2021}
Isaac Caswell, Julia Kreutzer, Lisa Wang, Ahsan Wahab, Daan van Esch,
  Nasanbayar Ulzii-Orshikh, Allahsera Tapo, Nishant Subramani, Artem Sokolov,
  Claytone Sikasote, Monang Setyawan, Supheakmungkol Sarin, Sokhar Samb,
  Benoît Sagot, Clara Rivera, Annette Rios, Isabel Papadimitriou, Salomey
  Osei, Pedro Javier~Ortiz Suárez, Iroro Orife, Kelechi Ogueji, Rubungo~Andre
  Niyongabo, Toan~Q. Nguyen, Mathias Müller, André Müller,
  Shamsuddeen~Hassan Muhammad, Nanda Muhammad, Ayanda Mnyakeni, Jamshidbek
  Mirzakhalov, Tapiwanashe Matangira, Colin Leong, Nze Lawson, Sneha Kudugunta,
  Yacine Jernite, Mathias Jenny, Orhan Firat, Bonaventure F.~P. Dossou, Sakhile
  Dlamini, Nisansa de~Silva, Sakine~Çabuk Ballı, Stella Biderman, Alessia
  Battisti, Ahmed Baruwa, Ankur Bapna, Pallavi Baljekar, Israel~Abebe Azime,
  Ayodele Awokoya, Duygu Ataman, Orevaoghene Ahia, Oghenefego Ahia, Sweta
  Agrawal, and Mofetoluwa Adeyemi.
\newblock Quality at a {Glance}: {An} {Audit} of {Web}-{Crawled} {Multilingual}
  {Datasets}.
\newblock April 2021.
\newblock arXiv: 2103.12028.

\bibitem{convertino_usefulness_2018}
Irma Convertino, Sara Ferraro, Corrado Blandizzi, and Marco Tuccori.
\newblock The usefulness of listening social media for pharmacovigilance
  purposes: a systematic review.
\newblock {\em Expert Opinion on Drug Safety}, 17(11):1081--1093, November
  2018.

\bibitem{lynx}
Thomas~E. Dickey.
\newblock Lynx - the text web-browser, 2021.
\newblock Accessed: 2021-09-02.

\bibitem{ding_jel_2021}
Wanying Ding, Vinay~K. Chaudhri, Naren Chittar, and Krihshna Konakanchi.
\newblock {JEL}: {Applying} {End}-to-{End} {Neural} {Entity} {Linking} in
  {JPMorgan} {Chase}.
\newblock {\em Proceedings of the AAAI Conference on Artificial Intelligence},
  35(17):15301--15308, May 2021.
\newblock Number: 17.

\bibitem{el-kishky_ccaligned_2020}
Ahmed El-Kishky, Vishrav Chaudhary, Francisco Guzmán, and Philipp Koehn.
\newblock {CCAligned}: {A} {Massive} {Collection} of {Cross}-{Lingual}
  {Web}-{Document} {Pairs}.
\newblock In {\em Proceedings of the 2020 {Conference} on {Empirical} {Methods}
  in {Natural} {Language} {Processing} ({EMNLP})}, pages 5960--5969, Online,
  November 2020. Association for Computational Linguistics.

\bibitem{fu_spanner_2021}
Jinlan Fu, Xuanjing Huang, and Pengfei Liu.
\newblock {SpanNER}: {Named} {Entity} {Re}-/{Recognition} as {Span}
  {Prediction}.
\newblock {\em arXiv:2106.00641 [cs]}, June 2021.
\newblock arXiv: 2106.00641.

\bibitem{selenium}
Jason Huggins, Paul Gross, and Jie~Tina Wang.
\newblock justext - heuristic based boilerplate removal tool, 2021.
\newblock Accessed: 2021-09-02.

\bibitem{ia-commons}
Ilya Kreymer, Sebastian Nagel, Andy Jackson, and Noah Levitt.
\newblock Iipc web archive commons - utility code for openwayback and other
  projects., 2021.
\newblock Accessed: 2021-09-02.

\bibitem{lang_textsweeper_2012}
Heinz-Peter Lang, Gerhard Wohlgenannt, and Albert Weichselbraun.
\newblock {TextSweeper} - {A} {System} for {Content} {Extraction} and
  {Overview} {Page} {Detection}.
\newblock In {\em International {Conference} on {Information} {Resources}
  {Management} ({Conf}-{IRM})}, Vienna, Austria, 2012. AIS.

\bibitem{li_effect_2014}
Qing Li, TieJun Wang, Ping Li, Ling Liu, Qixu Gong, and Yuanzhu Chen.
\newblock The effect of news and public mood on stock movements.
\newblock {\em Information Sciences}, 278:826--840, September 2014.

\bibitem{mikolov_distributed_2013}
Tomas Mikolov, Ilya Sutskever, Kai Chen, Gregory~S. Corrado, and Jeffrey Dean.
\newblock Distributed {Representations} of {Words} and {Phrases} and their
  {Compositionality}.
\newblock In {\em Advances in {Neural} {Information} {Processing} {Systems} 26:
  27th {Annual} {Conference} on {Neural} {Information} {Processing} {Systems}
  2013. {Proceedings} of a meeting held {December} 5-8, 2013, {Lake} {Tahoe},
  {Nevada}, {United} {States}}, pages 3111--3119, 2013.

\bibitem{cheerio}
Matthew Mueller, Felix Böhm, Jugglin Mike, and David Chambers.
\newblock Cheerio - fast, flexible, and lean implementation of core jquery
  designed specifically for the server., 2021.
\newblock Accessed: 2021-09-02.

\bibitem{doccano}
Hiroki Nakayama, Takahiro Kubo, Junya Kamura, Yasufumi Taniguchi, and Xu~Liang.
\newblock {doccano}: Text annotation tool for human, 2018.
\newblock Accessed: 2021-09-02.

\bibitem{pennington_glove:_2014}
Jeffrey Pennington, Richard Socher, and Christopher Manning.
\newblock Glove: {Global} {Vectors} for {Word} {Representation}.
\newblock In {\em Proceedings of the 2014 {Conference} on {Empirical} {Methods}
  in {Natural} {Language} {Processing} ({EMNLP})}, pages 1532--1543, Doha,
  Qatar, October 2014. Association for Computational Linguistics.

\bibitem{peters_content_2013}
Matthew~E. Peters and Dan Lecocq.
\newblock Content extraction using diverse feature sets.
\newblock pages 89--90. ACM, May 2013.

\bibitem{reis_transformers_2021}
Eduardo Souza~Dos Reis, Cristiano André~Da Costa, Diórgenes Eugênio~Da
  Silveira, Rodrigo~Simon Bavaresco, Rodrigo Da~Rosa Righi, Jorge
  Luis~Victória Barbosa, Rodolfo~Stoffel Antunes, Márcio~Miguel Gomes, and
  Gustavo Federizzi.
\newblock Transformers aftermath: current research and rising trends.
\newblock {\em Communications of the ACM}, 64(4):154--163, March 2021.

\bibitem{beautifulsoup}
Leonard Richardson.
\newblock Beautiful soup - a library that makes it easy to scrape information
  from web pages, 2021.
\newblock Accessed: 2021-09-02.

\bibitem{boilerpy3}
John Riebold.
\newblock Boilerpy3 - python port of boilerpipe library, 2021.
\newblock Accessed: 2021-09-02.

\bibitem{scharl_semantic_2017}
Arno Scharl, David Herring, Walter Rafelsberger, Alexander Hubmann-Haidvogel,
  Ruslan Kamolov, Daniel Fischl, Michael Föls, and Albert Weichselbraun.
\newblock Semantic {Systems} and {Visual} {Tools} to {Support} {Environmental}
  {Communication}.
\newblock {\em IEEE Systems Journal}, 11(2):762--771, 2017.

\bibitem{suarez_asynchronous_2019}
Pedro Javier~Ortiz Suárez, Benoît Sagot, and Laurent Romary.
\newblock Asynchronous {Pipeline} for {Processing} {Huge} {Corpora} on {Medium}
  to {Low} {Resource} {Infrastructures}.
\newblock Leibniz-Institut für Deutsche Sprache, July 2019.

\bibitem{html2text}
Aaron Swartz.
\newblock html2text - a python script that converts a page of html into clean,
  easy-to-read plain ascii text, 2021.
\newblock Accessed: 2021-09-02.

\bibitem{wang_review_2020}
Zhaoxia Wang, Seng-Beng Ho, and Erik Cambria.
\newblock A review of emotion sensing: categorization models and algorithms.
\newblock {\em Multimedia Tools and Applications}, January 2020.

\bibitem{weichselbraun_harvest_2020}
Albert Weichselbraun, Adrian M.~P. Brasoveanu, Roger Waldvogel, and Fabian
  Odoni.
\newblock Harvest - {An} {Open} {Source} {Toolkit} for {Extracting} {Posts} and
  {Post} {Metadata} from {Web} {Forums}.
\newblock In {\em 2020 {IEEE}/{WIC}/{ACM} {International} {Joint} {Conference}
  on {Web} {Intelligence} and {Intelligent} {Agent} {Technology} ({WI}-{IAT})},
  pages 438--444, December 2020.

\bibitem{weichselbraun_adapting_2021}
Albert Weichselbraun, Philipp Kuntschik, Vincenzo Fancolino, Mirco Saner, and
  Vinzenz Wyss.
\newblock Adapting {Data}-{Driven} {Research} to the {Fields} of {Social}
  {Sciences} and the {Humanities}.
\newblock {\em Future Internet}, 13(3), 2021.
\newblock Accepted 22 February 2021.

\bibitem{xue_mt5_2021}
Linting Xue, Noah Constant, Adam Roberts, Mihir Kale, Rami Al-Rfou, Aditya
  Siddhant, Aditya Barua, and Colin Raffel.
\newblock {mT5}: {A} {Massively} {Multilingual} {Pre}-trained {Text}-to-{Text}
  {Transformer}.
\newblock In {\em Proceedings of the 2021 {Conference} of the {North}
  {American} {Chapter} of the {Association} for {Computational} {Linguistics}:
  {Human} {Language} {Technologies}}, pages 483--498, Online, June 2021.
  Association for Computational Linguistics.

\end{thebibliography}

\end{document}